\documentclass[ onecolumn, 08 pt,amsmath,amssymb]{revtex4-2}

\usepackage[titletoc]{appendix}
\usepackage[english]{babel}
\usepackage{mathtools}
\usepackage{tabularx}
\allowdisplaybreaks
\usepackage{multirow}
\usepackage{comment}
\usepackage{graphicx}
\usepackage{graphics}
\usepackage{amsmath}
\usepackage{color}
\usepackage{appendix}
\usepackage{float}
\begin{document}
\title{Study of tetraquarks in dipole-dipole interaction potential}
\author{Sindhu D G}\email{sindhudgdarbe@gmail.com}   \author{Akhilesh Ranjan}\email{ak.ranjan@manipal.edu}
\affiliation{Department of Physics \\ Manipal Institute of Technology, \\Manipal University, \\ Udupi, 576104, India}
\begin{abstract}
In recent years tetraquark and pentaquark states have received much attention 
due to the significant experimental findings. In this work masses of some heavy 
tetraquarks are estimated by considering the spin-spin interaction, the 
dipole-dipole interaction, and the meson-meson interaction. It is found that 
such interactions give significant mass contributions. The Regge trajectory of X(3872) state is also studied and is found to be non-linear. 
Masses of some tetraquark states are also proposed. \end{abstract}

\maketitle
\section{INTRODUCTION}
\noindent In nature, the most elementary particles (experimentally observed) 
are quarks, leptons, and gauge particles. The quarks carry color 
degree of freedom. Because of color confinement one can not 
observe a free quark in nature. All forms of quark matter in 
nature are color singlet. All these quark matter around us are 
made of mesons (quark - antiquartk) and baryons (three quarks). In principle any combination of color singlet multiquark systems 
can exist in nature. Therefore, search for multiquark systems beyond baryons has always been an area of interest for many physicists \cite{guidry,cheng}. First time the existence of these particles was proposed by Gellmann \cite{gellmann}. In early days because of limited experimental as well as computational facilities, many experimental results on tetraquarks became inconclusive. Recently many experimental findings have confirmed the existence of tetraquarks. Tetraquarks are the hadrons containing two quarks and two antiquarks. Since these are very short lifetime particles therefore, their many properties are still unknown. Study of tetraquarks is very important because it helps us to understand the physics of higher order quark matter, nature of color confinement, and strong forces. 

In this work we shall consider tetraquarks containing atleast one heavy quark. 
However, in popular compositions, its quark contents are $qQ\bar{q}\bar{Q}$, 
where 
$q$ represents a light (up, down
or strange) quark, $Q$ represents a heavy (charm or bottom) quark, and 
$\bar{q}, \bar{Q}$ are the corresponding antiquarks \cite{Zyla}. 
But the QCD based studies also indicate the presence of tetraquarks with 
fully heavy flavors \cite{hughes,chen}.  
Because of their large mass and very short lifetime, tetraquarks remained 
undetected till the last century.
In 2003 the Belle experiment in Japan discovered a tetraquark state 
$X$(3872)\cite{choi}, later confirmed by BABAR Collaboration \cite{aubert}, 
the Collider Detector at Fermilab experimental collaboration (CDF II)
\cite{acosta,aaltonen}, D0 experiment\cite{abazov}, LHCb (Large Hadron 
Collider beauty)\cite{aaij1} and CMS (Compact Muon Solenoid) etc 
\cite{chatrchyan}. Following its discovery, many new tetraquark candidates 
have also been discovered. 
Later other tetraquark states Y(4260) at BABAR in 2005 \cite{ablikim}, Z(3900) 
at BESIII in 2013 \cite{aaij2} were also discovered. Recently the LHCb 
collaboration observed a new tetraquark state with all heavy quark flavors 
\cite{aaij3}.  
So  with improved experimental facilities many new tetraquarks 
states are discovered and the quest is still on.
Because of the lack of sufficient experimental data, the 
nature of these states is still unclear and the work is under progress.

Some progress is also reported in lattice QCD approach, which is actually a  
computational technique on tetraquarks. In a work by Prelovsek 
and Leskovec it is argued that X(3872) is a bound state pole of $DD^*$ 
scattering \cite{prelovsek}. In a separate work HAL QCD collaboration have 
studied the $Z_c(3900)$ by the method of coupled-channel scattering in lattice 
QCD and shown that it is a threshold cusp \cite{hal}. In another work Francis 
{\it et al} have studied the possibility of $qq'\bar{b}\bar{b}$ in 
lattice QCD framework and have shown the existence of strong-interaction-stable 
$J^P = 1^+$ tetraquarks \cite{francis}. Bicudo, Scheunert, and Wagner have 
studied spin effects in $qq\bar{b}\bar{b}$ like systems using the 
Born-Oppenheimer approximation. In this work they have predicted the 
existence of $ud\bar{b}\bar{b}$ tetraquark system with quantum 
number $I(J^P) = 0(1^+)$ \cite{bicudo}. Junnarkar, Mathur, and Padmanath have 
done a lattice QCD study and estimated mass and energy levels of 
$qq\bar{Q}\bar{Q}$ type of tetraquark systems \cite{junnarkar}. 
This way we find that lattice QCD results also support the existence of 
tetraquarks and are helpful to explain their properties.

At present we do not have a theory to study tetraquark systems. Therefore, it 
is studied with some models. The properties of tetraquarks are studied by 
interpreting their internal structure as cusps, hadron molecules, 
diquarks-antidiquarks etc, using suitable potential models \cite{ali,yrliu}. 
Ebert, Faustov, and 
Galkin have studied the mass spectra of hidden charm and bottom quark 
containing tetraquarks in diquark-antidiquark picture. They have shown that 
the $X(3872)$ can be interpreted as hidden charm tetraquark \cite{ebert}. 
In a work Liu \textit{et al}, have studied the open charm decay modes of state 
Y(4630) by considering tetraquark with a bound state of a diquark and an 
anti-diquark \cite{liu}. Gutsche, Kesenheimer, and  Lyubovitskij1 have studied 
the $Z^{+}_{c}$(3900) tetraquarks by considering it as a hadronic molecule 
and interpreted it's decay widths \cite{gutsche}. Rathaud and Rai have studied 
some heavy quarks systems as di-hadronic molecules and shown that there 
should be dipole like interaction between two color neutral states 
\cite{rathaud}. Mass and decay widths of $bb\bar{b}\bar{b}$ was estimated by 
Esposito and Polosa using diquark-antidiquark interaction model 
\cite{esposito}. In a separate work the mass spectrum of $c\bar{c}c\bar{c}$ 
states are also determined using the dynamical diquark model by Giron and 
Lebed \cite{giron}. Recently Cheng {\it et al} have estimated the mass spectrum 
and constraints of double-heavy tetraquark states with heavy 
diquark-antidiquark symmetry and the chromomagnetic interaction \cite{yao}. 
Hence we see that various models are used to mainly study the mass spectrum 
and decay properties of tetraquarks for different tetraquark systems. 

In order to study the properties of highly confined quarks, different models 
are used. The string model of hadrons is one such model which gives the
Regge trajectories of hadrons. The study of Regge trajectory is important 
because it not only gives estimation of hadron masses, also directly correlates 
mass with the angular momentum of a hadron 
\cite{guidry,cheng}. It is shown that the Regge trajectories for mesons, 
baryons, and pentaquarks are non-linear 
\cite{olsson,burakovsky,hothi,zhu,nandan1,nandan2}.

Various theoretical models have been developed to study the Regge trajectories 
of hadrons. Some of the theoretical models are, string models such as Olsson 
model \cite{olsson}, Soloviev string quark model \cite{soloviev}, 
non-relativistic quark models such as Inopin model \cite{inopin1,inopin2}, 
Martin model \cite{martin}, relativistic and semi-relativistic models like 
Basdevant model \cite{basdevant}, Semay model \cite{semay1,semay2} etc. 

In this paper we have estimated masses of different heavy tetraquarks 
by considering the spin-spin interaction, meson-meson interaction, as well as 
dipole-dipole interactions. We have found that these interactions 
cause significant corrections. At the end we have also discussed about the 
Regge trajectory of X(3872) state. 
\section{Formulation}
\label{sec:2}
The masses of different tetraquark states are determined by considering 
different interactions. Different interactions will have different 
contribution. In this work the spin-spin, diquark-antidiquark and dipole 
interactions are considered. Initially the masses are determined using 
constituent 
quark model which contains the individual quark masses and spin-spin 
interactions. A tetraquark system $q_1q_2\bar{q_3}\bar{q_4}$ 
($q = u,d,s,c$ quarks) can have three types of interactions namely, 
diquark-antidiquark interaction 
$[qq]$[$\bar{q}\bar{q}]$, dipole-dipole or meson-meson interactions 
$[q\bar{q}]$[$q\bar{q}]$. Here the dipole interaction can be color as well as 
electromagnetic. Here the electromagnetic interaction between dipoles is 
considered and it is assumed that the dipoles formed are sufficiently far 
from each other so the color interaction can be ignored. Therefore, the total 
mass of the tetraquark system is
\[M_{th} = \sum\limits_{i=1}^{4}m_{i}+E_{spin}+\triangle M\]
where, $\triangle M = (E_{diquark}+E_{meson}+E_{dipole})_{avg}$,
$E_{spin}$ is the spin-spin interaction energy, $E_{diquark}$ is the 
diquark-antidiquark interaction energy, $E_{meson}$ is the meson-meson 
interaction energy and $E_{dipole}$ is the dipole-dipole 
interaction energy. Let us discuss these contributions separately. 
\subsection{Spin-spin interaction}
\label{subsec:1}
The total mass is due to the constituent quark masses as well as their 
spin-spin interactions. The standard spin-spin interaction term for quarks 
can be written as 
\[E_{spin}=A^{\prime}\sum_{\substack{j\neq k \\ j,k=1}}^4 \left(\frac{S_{j}\cdot S_{k}}{m_{j}m_{k}}\right)\]
Let
\[M_{spin}=\sum\limits_{i}m_{i}+E_{spin}.\]
The summation is over constituent quarks. $A^{\prime}$ is a constant. Here 
$A^{\prime}$ is a parameter kept by hand and is taken to be equal to 
$A^{\prime}=2.36\times 10^{7}MeV^{3}$ (for tetraquark with charm quark) and 
$A^{\prime}=-1.6\times 10^{8}MeV^{3}$ (for tetraquark with bottom quark). 
If $S_{jk} = S_j + S_k$ then 
\[S_{j}\cdot S_{k}=\frac{S^{2}_{jk}-S^{2}_{j}-S^{2}_{k}}{2}\]  $j,k$ are the 
different flavor constituents of the tetraquark.\\
Here the constituent quark masses are $m_{u}=336MeV$, $m_{d}=340MeV$, 
$m_{s}=486MeV$, $m_{c}=1550MeV$, and $m_b=4730MeV$ \cite{scadron}. The masses 
due to spin-spin interaction is shown in table 2. 
 There will be actually many type of interactions among quarks/antiquarks. 
 Based on their contribution we consider only dominating terms. Here 
 the diquark-antidiquark, meson-meson interactions are considered. Actually 
 in the meson-meson interaction it is found 
 that the color interaction contribution is smaller than the electric 
 dipole-dipole interaction.

\subsection{Diquark-antidiquark interaction}
\label{subsec:2}
A diquark is a two quark composition forming a bound state. The formation of 
bound state is mainly due to one-gluon exchange potential \cite{debastiani}. 
According to $SU(3)$, diquarks can form color triplet or color sextet with 
color factor $-\frac{2}{3}$ (triplet) and $\frac{1}{3}$ (sextet).  
Diquark-antidiquark systems are considered as two-body systems and the 
potential chosen in 
this case is Cornell-like potential. The sextet color configuration is 
considered 
because only then the potential will show confining nature.
\[V_{Cornell}=ar-k_{s}\frac{b}{r}\]
This potential is solved by Kuchin \textit{et al}, \cite{kuchin} using the 
Schroedinger's equation, and the expression for energy is,
\[E_{diquark}=\frac{3a}{\delta}-\frac{\left(k_{s}b+\frac{3a}{\delta^{2}}\right)^{2}2\mu}{\left((2n+1)\pm\sqrt{1+4l(l+1)
+\frac{8\mu a}{\delta^{3}}}\right)^{2}}\]
Here $a$, $b$, and $\delta$ are constants, $a=0.2GeV^{2}$, $b=1.244$ and 
$\delta=0.231GeV$ (for charm quark) and $b=1.569$ and $\delta=0.378GeV$ (for bottom quark) \cite{kuchin} and $k_{s}=\frac{1}{3}$. In calculation 
we have chosen the negative terms only because the positive term causes 
unacceptable errors. 
\subsection{Meson-meson interaction (using Cornell potential)}
\label{subsec:3}
Just like diquarks, quark, and antiquarks also interact via one-gluon exchange 
potential. The color factor depends on the color state of the quarks. It can 
form a color singlet or octet. The color factor can take values $-\frac{4}{3}$ 
(singlet) or $\frac{1}{6}$ (octet). The octet-octet interactions are considered 
to include the color interaction. The energy due to mesonic 
interactions are determined using the diquark energy expression.
\subsection{Meson-meson interaction (using Yukawa potential)}
\label{subsec:4}
In this section the Yukawa potential of the following form is considered,
\[V(r)=-V_{0}\frac{e^{-ar}}{r}\]
where, $V_{0}=0.5$. Hamzavi {\it et al} had obtained the solution of the 
Yukawa potential as \cite{hamzavi},
\[E_{nl}=-\frac{a^{2}}{2m}\frac{\left(\frac{mV_{0}}{a}-(n+1)^{2}-l(2n+l+2)\right)^{2}}{(n+l+1)^{2}}\]
In this work $n=0$, $l=0$ and $a=0.686 fm^{-1}$.
\subsection{Dipole-dipole interaction}
\label{subsec:5}
Here the tetraquark is considered as the two dipoles interacting.
\begin{figure}
\resizebox{0.35\textwidth}{!}
{\includegraphics{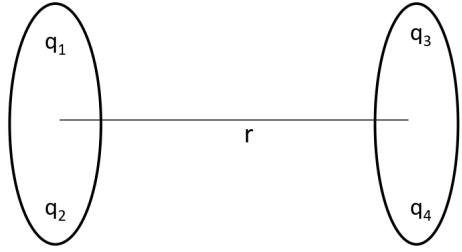}}
\caption{Dipole-dipole interaction}
\label{fig:1}      
\end{figure}
Let us consider a tetraquark as two electromagnetically interacting dipoles. 
It is assumed that the separation between dipoles (inside a tetraquark) is 
 large so the color interaction can be ignored. The dipoles may interact 
 with all orientations due to thermal effect. Hence it must should be thermally 
 averaged. The thermally averaged dipole-dipole interaction potential is given 
 by,
\[V(r)=-\frac{C}{r^{6}}\]
where, 
\[C=\frac{2}{3}\frac{\mu^{2}_{A}\mu^{2}_{B}}{(4\pi \epsilon_{0})^{2}}\frac{1}{k_{B}T}\]
$\mu_{A}$ and $\mu_{B}$ are the dipole moment of two dipoles and $r$ is the 
distance between the center of two dipoles.\\
According to Gao \cite{gao} on solving the radial Schroedinger's equation for 
$-\frac{C}{r^{6}}$ potential, the expression for energy can be determined 
\cite{gao}.
\[ E=\Delta \times 16 \times \frac{\hbar^{2}}{2\mu}\frac{1}{\beta^{2}}\]
Here, \[\beta=\left(\frac{2\mu C}{\hbar^{2}}\right)^{\frac{1}{4}}\] $\Delta$ is 
the critical scaled energy and $\mu$ is the reduced mass of the dipoles. 
The value of $\Delta$ for different angular momentum states are shown in 
Table \ref{tab:1}. The simplified expression for energy in (MeV) can be expressed as, 
\begin{eqnarray}
\label{eqn-1}
E=3.7\times 10^{7} \sqrt{\frac{k_{B}T}{\mu^{3}}}\frac{\Delta}{f_{A}f_{B}d_{A}d_{B}}
\end{eqnarray} 
Here, $T$ and $\mu$ are expressed in MeV and $d_{A}$ and $d_{B}$ are the dipole 
lengths expressed in $fm$. The critical scaled energy for different angular 
momenta $l$ is given in the following table. Here the temperature is at 
$T = 0.1 MeV$. We have chosen the low temperature because we are interested in 
ground state energy or lowest mass.

\begin{table}[H]
\centering
\caption{Critical scaled energy for different angular momenta}
\label{tab:1}      
\begin{tabular}{lll}
\hline\noalign{\smallskip}
$l$&&$\Delta$\\
\noalign{\smallskip}\hline\noalign{\smallskip}
$s$&&9.654418$\times 10^{-2}$\\
$p$&&1.473792$\times 10^{-1}$\\
$d$&&4.306921$\times 10^{-1}$\\
$f$&&1.580826\\
$g$&&2.073296\\
\noalign{\smallskip}\hline
\end{tabular}
\end{table}
The energy corrections due to diquark-antiquark, meson-meson (using Cornell 
potential) and dipole-dipole interactions are determined and are given in 
Table \ref{tab:2}.\\
The energy corrections due to diquark-antiquark, meson-meson (using Yukawa 
potential) and dipole-dipole interactions are determined and are given in Table 
\ref{tab:3}.\\
\begin{table}[H]
\centering
\caption{Energy corrections due to diquark, meson (using Cornell potential) and dipole interactions}
\label{tab:2} 
\begin{tabular}{|c|c|c|cc|cc|c|c|c|}
\hline \noalign{\smallskip} 
State&$M_{spin}$&$E_{diquark}$&&$E_{meson}$&&$E_{dipole}$&$\triangle M$&$M_{th}$&\%\\
&&&&&&&&&error\\
(Exp)&(MeV)&(MeV)&&(MeV)&&(MeV)&(MeV)&(MeV)&\\
\noalign{\smallskip}
\hline
&&&&&&&&&\\
$\bar{c}ud\bar{s}$&&$ud-\bar{c}\bar{s}$&$\bar{c}u-d\bar{s}$&$\bar{c}d-u\bar{s}$&$c\bar{u}-d\bar{s}$&$\bar{c}\bar{s}-ud$&&&\\
$X_{0}(2900)$&2558&-81
&51
&50
&367&392&260
&2818&2.8\\
&&&&&&&&&\\
\hline
&&&&&&&&&\\
$c\bar{c}u\bar{u}$&&$cu-\bar{c}\bar{u}$&$c\bar{c}-u\bar{u}$&$c\bar{u}-\bar{c}u$&$c\bar{c}-u\bar{u}$&$c\bar{u}-\bar{c}u$&&&\\
$X(3872)$&3672&88
&177
&39
&194&87&195
&3867&0.13\\
&&&&&&&&&\\
\hline
&&&&&&&&&\\
$uc\bar{d}\bar{c}$&&$uc-\bar{d}\bar{c}$&$u\bar{d}-c\bar{c}$&$u\bar{c}-c\bar{d}$&&&&&\\
$Z^{+}_{c}(3900)$&3677&88
&41
&177
&&
&153&3830&1.8\\	
&&&&&&&&&\\
\hline
&&&&&&&&&\\
$cs\bar{c}\bar{s}$&&$cs-\bar{c}\bar{s}$&$c\bar{c}-s\bar{s}$&$c\bar{s}-s\bar{c}$&$c\bar{c}-s\bar{s}$&$cs-\bar{c}\bar{s}$&&&\\
$Y(4140)$&4013&105
&193
&120
&250&138&269
&4282&3.4\\
&&&&&&&&&\\
\hline
&&&&&&&&&\\
$d\bar{u}\bar{c}c$&&$dc-\bar{u}\bar{c}$&$d\bar{u}-\bar{c}c$&$d\bar{c}-\bar{u}c$&&&&&\\
$Z(4430)$&3677&88
&41
&177
&&&153
&3830&13.5\\	
&&&&&&&&&\\
\hline
&&&&&&&&&\\
$cs\bar{c}\bar{s}$&&$cs-\bar{c}\bar{s}$&$c\bar{c}-s\bar{s}$&$c\bar{s}-s\bar{c}$&$c\bar{c}-s\bar{s}$&$cs-\bar{c}\bar{s}$&&&\\
$X(4700)$&4013&105
&193
&120
&250&138&269
&4282&8.8\\
&&&&&&&&&\\
\hline
&&&&&&&&&\\                                                                     $cc\bar{c}\bar{c}$&&$cc-\bar{c}\bar{c}$&$c\bar{c}-c\bar{c}$&$c\bar{c}-c\bar{c}$&$c\bar{c}-c\bar{c}$&$c\bar{c}-c\bar{c}$&&&\\
(6900)&6185&186
&271
&271
&41&41&270
&6455&6.4\\
&&&&&&&&&\\
\hline
\end{tabular}
\end{table}
\begin{table}[H]
\centering
\caption{Energy corrections due to spin-spin, diquark, meson (using Yukawa potential) and dipole interactions}
\label{tab:3} 
\begin{tabular}{|c|c|c|cc|cc|c|c|c|}
\hline \noalign{\smallskip} 
State&$M_{spin}$&$E_{diquark}$&&$E_{meson}$&&$E_{dipole}$&$\triangle M$&$M_{th}$&\%\\
&&&&&&&&&\\
(Exp)&(MeV)&(MeV)&&(MeV)&&(MeV)&(MeV)&(MeV)&error \\
\noalign{\smallskip}
\hline
&&&&&&&&&\\
$\bar{c}ud\bar{s}$&&$ud-\bar{c}\bar{s}$&$\bar{c}u-d\bar{s}$&$\bar{c}d-u\bar{s}$&$c\bar{u}-d\bar{s}$&$\bar{c}\bar{s}-ud$&&&\\
$X_0(2900)$&2558&-81
&-5&-5&367&392&223&2781&4.1\\
&&&&&&&&&\\
\hline
&&&&&&&&&\\
$c\bar{c}u\bar{u}$&&$cu-\bar{c}\bar{u}$&$c\bar{c}-u\bar{u}$&$c\bar{u}-\bar{c}u$&$c\bar{c}-u\bar{u}$&$c\bar{u}-\bar{c}u$&&&\\
$X(3872)$&3672&88
&-18&-59&194&87&97&3769&2.6\\
&&&&&&&&&\\
\hline
&&&&&&&&&\\
$uc\bar{d}\bar{c}$&&$uc-\bar{d}\bar{c}$&$u\bar{d}-c\bar{c}$&$u\bar{c}-c\bar{d}$&&&&&\\
$Z^+_c(3900)$&3677&88
&-17&-59&&&4&3681&5.6\\	
&&&&&&&&&\\
\hline
&&&&&&&&&\\
$cs\bar{c}\bar{s}$&&$cs-\bar{c}\bar{s}$&$c\bar{c}-s\bar{s}$&$c\bar{s}-s\bar{c}$&$c\bar{c}-s\bar{s}$&$cs-\bar{c}\bar{s}$&&&\\
$Y(4140)$&4013&105
&-40&-65&249&138&129&4142&0.05\\	
&&&&&&&&&\\
\hline
&&&&&&&&&\\
$d\bar{u}\bar{c}c$&&$dc-\bar{u}\bar{c}$&$d\bar{u}-\bar{c}c$&$d\bar{c}-\bar{u}c$&&&&&\\
$Z(4430)$&3677&88
&-17&-59&&&4&3681&17\\	
&&&&&&&&&\\
\hline
&&&&&&&&&\\
$cs\bar{c}\bar{s}$&&$cs-\bar{c}\bar{s}$&$c\bar{c}-s\bar{s}$&$c\bar{s}-s\bar{c}$&$c\bar{c}-s\bar{s}$&$cs-\bar{c}\bar{s}$&&&\\
$X(4700)$&4013&105
&-40&-65&249&138&129&4142&12\\
&&&&&&&&&\\
\hline
&&&&&&&&&\\
$cc\bar{c}\bar{c}$&&$cc-\bar{c}\bar{c}$&$c\bar{c}-c\bar{c}$&$c\bar{c}-c\bar{c}$&$c\bar{c}-c\bar{c}$&$c\bar{c}-c\bar{c}$&&&\\
(6900)&6185&185
&-132&-132&41&41&1&6186&10\\
&&&&&&&&&\\
\hline
\end{tabular}
\end{table}
\begin{table}[H]
\centering
\caption{Predicted mass spectra of some tetraquark states}
\label{tab:4} 
\begin{tabular}{|c|c|c|cc|cc|c|c|}
\hline \noalign{\smallskip} 
State&$M_{spin}$&$E_{diquark}$&&$E_{meson}$&&$E_{dipole}$&$\triangle M$&$M_{th}$\\
&&&&&&&&\\
(Exp)&(MeV)&(MeV)&&(MeV)&&(MeV)&(MeV)&(MeV)\\
\noalign{\smallskip}
\hline
&&&&&&&&\\
$ss\bar{s}\bar{s}$&&$ss-\bar{s}\bar{s}$&$s\bar{s}-s\bar{s}$&$s\bar{s}-s\bar{s}$&$s\bar{s}-s\bar{s}$&$s\bar{s}-s\bar{s}$&&\\
&2960&-952
&-678&-678&941&941&-142&2818\\
&&&&&&&&\\
\hline
&&&&&&&&\\
$uu\bar{d}\bar{b}$&&$uu-\bar{d}\bar{b}$&$u\bar{d}-u\bar{b}$&$u\bar{b}-u\bar{d}$&&&&\\
&6872&-831&-568&-568&&&-984&5888\\	
&&&&&&&&\\
\hline
&&&&&&&&\\
$dd\bar{u}\bar{b}$&&$dd-\bar{u}\bar{b}$&$d\bar{u}-d\bar{b}$&$d\bar{u}-d\bar{b}$&&&&\\
&6867&-826
&-568&-568&&&-981&5886\\	
&&&&&&&&\\
\hline
&&&&&&&&\\
$\bar{b}ud\bar{s}$&&$ud-\bar{b}\bar{s}$&$\bar{b}u-d\bar{s}$&$\bar{b}d-u\bar{s}$&&&&\\
&6797&-826
&-492&-488&&&-903&5894\\
&&&&&&&&\\
\hline
&&&&&&&&\\
$us\bar{s}\bar{b}$&&$us-\bar{s}\bar{b}$&$u\bar{s}-s\bar{b}$&$s\bar{s}-u\bar{b}$&&&&\\
&6758&-737
&-486&-428&&&-826&5932\\
&&&&&&&&\\
\hline
&&&&&&&&\\
$uu\bar{s}\bar{b}$&&$uu-\bar{s}\bar{b}$&$u\bar{s}-u\bar{b}$&$u\bar{b}-u\bar{s}$&&&&\\
&6800&-830
&-488&-488&&&-602&6198\\
&&&&&&&&\\
\hline
&&&&&&&&\\
$ss\bar{s}\bar{b}$&&$ss-\bar{s}\bar{b}$&$s\bar{s}-s\bar{b}$&$s\bar{s}-s\bar{b}$&$s\bar{s}-s\bar{b}$&$s\bar{s}-s\bar{b}$&&\\
&6748&-669
&-426&-426&430&430&-220&6528\\
&&&&&&&&\\
\hline
&&&&&&&&\\
$bb\bar{b}\bar{b}$&&$bb-\bar{b}\bar{b}$&$b\bar{b}-b\bar{b}$&$b\bar{b}-b\bar{b}$&$b\bar{b}-b\bar{b}$&$b\bar{b}-b\bar{b}$&&\\
&18931&-197
&-5&-5&30&30&-49&18882\\
&&&&&&&&\\
\hline
\end{tabular}
\end{table}
\begin{table}[H]
\centering
\caption{ Comparison between Cornell and
 Yukawa potential corrections}
\label{tab:5}  
\begin{tabular}{|c|c|c|}
\hline
&&\\
State&$M_{th}$&$M_{th}$\\
(Expt)&(Cornell)&(Yukawa)\\
(MeV)&(MeV)&(MeV)\\
\hline
$\bar{c}ud\bar{s}$&&\\
$X_{0}$(2900)&2818&2781\\
&&\\
\hline
$c\bar{c}u\bar{u}$&&\\
$X$(3872)&3867&3769\\
&&\\
\hline
$uc\bar{d}\bar{c}$&&\\
$Z_{c}^{+}$(3900)&3830&3681\\
&&\\
\hline
$cs\bar{c}\bar{s}$&&\\
$Y$(4140)&4282&4142\\
&&\\
\hline
$d\bar{u}\bar{c}c$&&\\
$Z$(4430)&3830&3681\\
&&\\
\hline
$cs\bar{c}\bar{s}$&&\\
$X$(4700)&4282&4142\\
&&\\
\hline
$cc\bar{c}\bar{c}$&&\\
(6900)&6455&6186\\
&&\\
\hline
\end{tabular}
\end{table}
\begin{table}[H]
\centering
\caption{Comparison of masses of tetraquarks, in units of MeV, calculated in 
this work and values from other different works.}
\label{tab:6} 
\begin{tabular}{ccccccccc}
\hline
&&&&&&&&\\
State&&&&present work&&&&other work\\
&&&&&&&&\\
\hline
&&&&&&&&\\
$ss\bar{s}\bar{s}$&&&&2818&&&&2200-2900\cite{feng}\\
&&&&&&&&\\
$uu\bar{d}\bar{b}$&&&&5888&&&&5977-6503\cite{qi}\\
&&&&&&&&\\
$dd\bar{u}\bar{b}$&&&&5886&&&&5977-6503\cite{qi}\\
&&&&&&&&\\
$\bar{b}ud\bar{s}$&&&&5894&&&&5584$\pm$137 \cite{agaev}\\
&&&&&&&&\\
$us\bar{s}\bar{b}$&&&&5932&&&&6388-6728\cite{qi}\\
&&&&&&&&\\
$uu\bar{s}\bar{b}$&&&&6198&&&&6194-6602\cite{qi}\\
&&&&&&&&\\
$ss\bar{s}\bar{b}$&&&&6528&&&&6682-6826\cite{qi}\\
&&&&&&&&\\
$bb\bar{b}\bar{b}$&&&&18882&&&&18690 \cite{yang}\\
&&&&&&&&\\
\hline
\end{tabular}
\end{table}
\section{Results and Discussion}
\label{sec:3}
In tetraquark mass calculation constituent quark masses, 
spin-spin, diquark-antidiquark, meson-meson, and dipole-dipole interactions 
are considered. From results 
it is clear that the main contribution is due to constituent quark masses 
and spin-spin interaction. The spin-spin interaction takes care of 
color as well as electromagnetic interactions simultaneously. Now for the 
diquark-antidiquark interactions the Cornell potential with 
appropriate color factors as mentioned earlier is chosen. 
In Table \ref{tab:2} the $E_{diquark}$ is calculated between diquark and antidiquark 
using the Cornell potential. In Table \ref{tab:3} the same is done with Yukawa potential. 
From tables \ref{tab:2} and \ref{tab:3} it is clear that the diquark-antidiquark interaction energy 
is very small compared to the meson-meson interaction where the 
tetraquark is considered as two interacting mesons. The decay modes of 
tetraquark also 
support this because most of the decay mode show that decay of tetraquark 
leads to atleast two mesons. On the other hand for diquark-antidiquark picture 
there should also be possibility of two baryons after tetraquark decay. In 
Table \ref{tab:5} it is 
also interesting to see that the mass corrections due to the Cornell potential 
is smaller than the Yukawa potential corrections. However, the results with the 
Cornell potential are much closer to the experimental values. So the color 
interaction picture appears to be more realistic. For tetraquarks 
two types of mesonic interactions are possible. Hence both the possibilities 
are considered with equal probability. Actually the probability of a particular 
configuration can be taken from the branching ratio of decay modes obtained 
from experiments. In meson-meson case dipole interactions 
are also possible. It is interesting to see that such interactions also cause 
significant mass corrections.

From the tables it is clear that for low mass tetraquarks the major 
contributions are coming due to the diquark-antidiquark, meson-meson, 
dipole-dipole, and spin-spin interactions. The difference from experimental 
results are very small. On the other hand if we go to larger mass states 
the difference gradually increases which tells that other effects will also 
be dominating. For $Z(4430)$ and $Z^+_c(3900)$ states the dipole interaction 
contribution is very small $(<MeV)$ so it is ignored.

In Table \ref{tab:4} the masses of some tetraquark states are proposed and also 
compared with other theoretical results in Table \ref{tab:6}. In this case  
only the Cornell potential for meson and diquark interactions is considered. 
Our results are in good agreement with other theoretical works. 

\begin{figure}[H]
\centering
\caption{The plot of energy correction vs temperature}
\resizebox{0.4\textwidth}{!}{%
\includegraphics{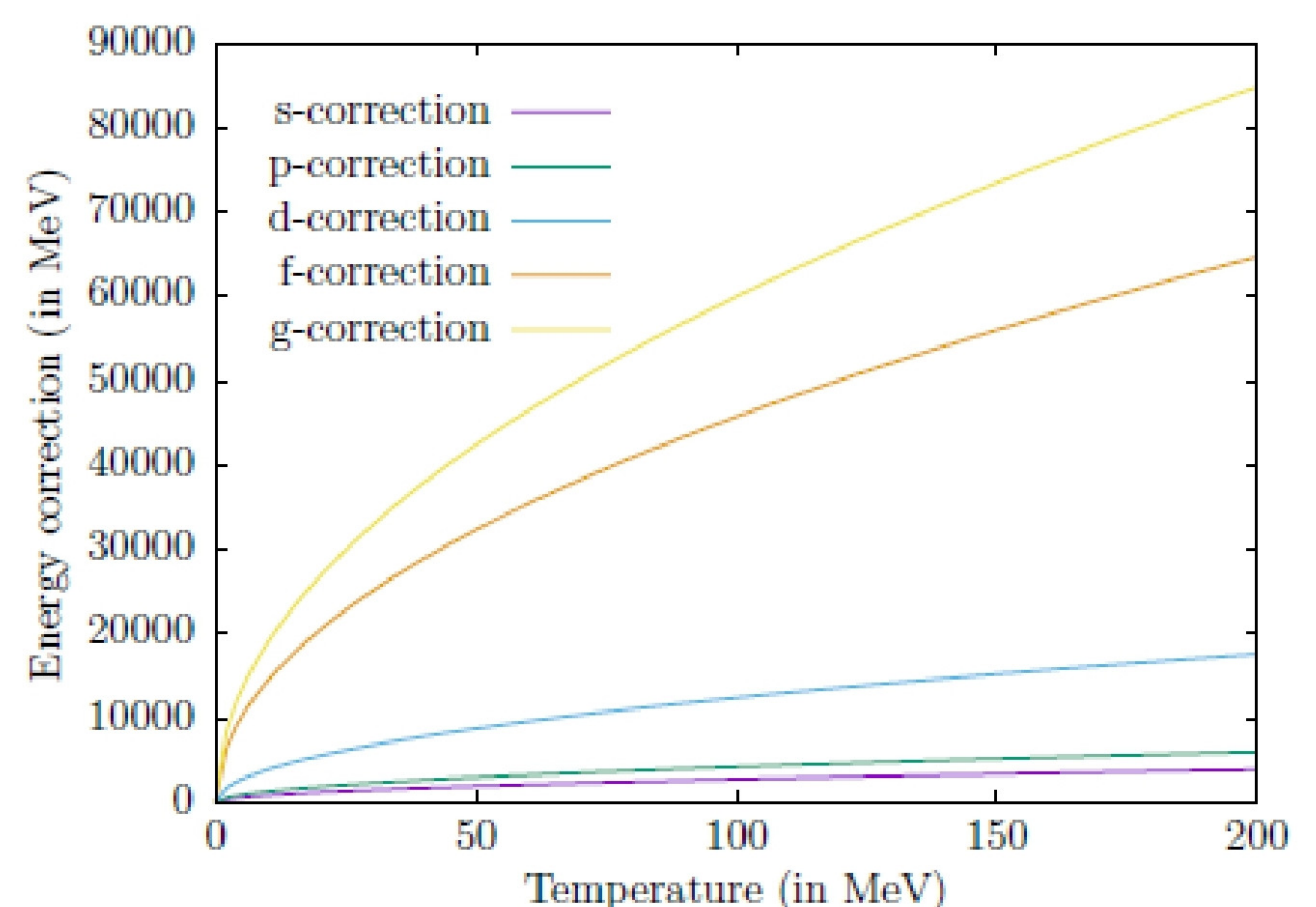}}
\label{fig:2}       
\end{figure}

\begin{figure}[H]
\centering
\caption{The plot of energy correction vs mass}
\resizebox{0.4\textwidth}{!}{%
  \includegraphics{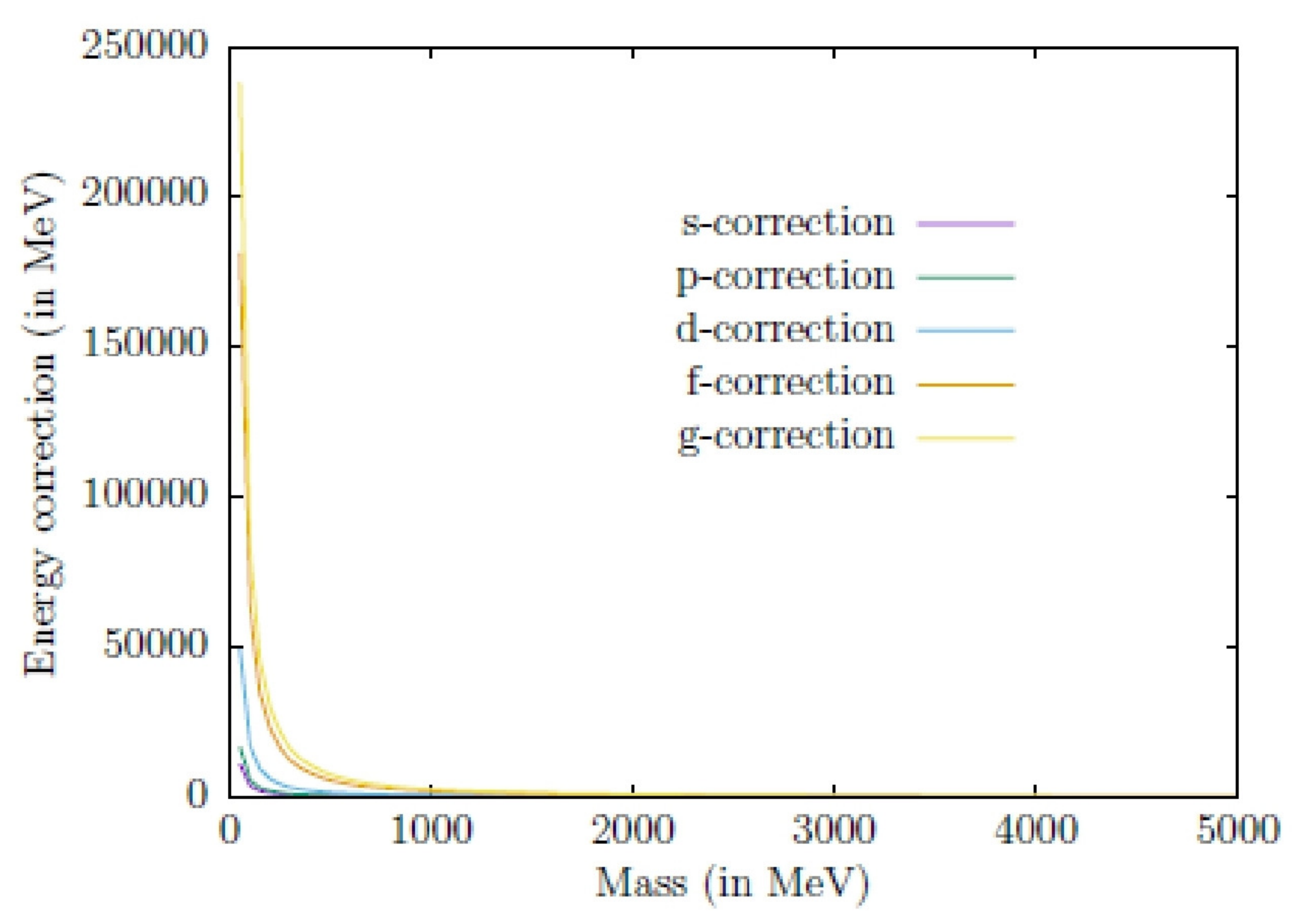}
}
\label{fig:3}       
\end{figure}

\begin{figure}[H]
\centering
\caption{Regge trajectory of $X$ (3872)}
\resizebox{0.4\textwidth}{!}{%
  \includegraphics{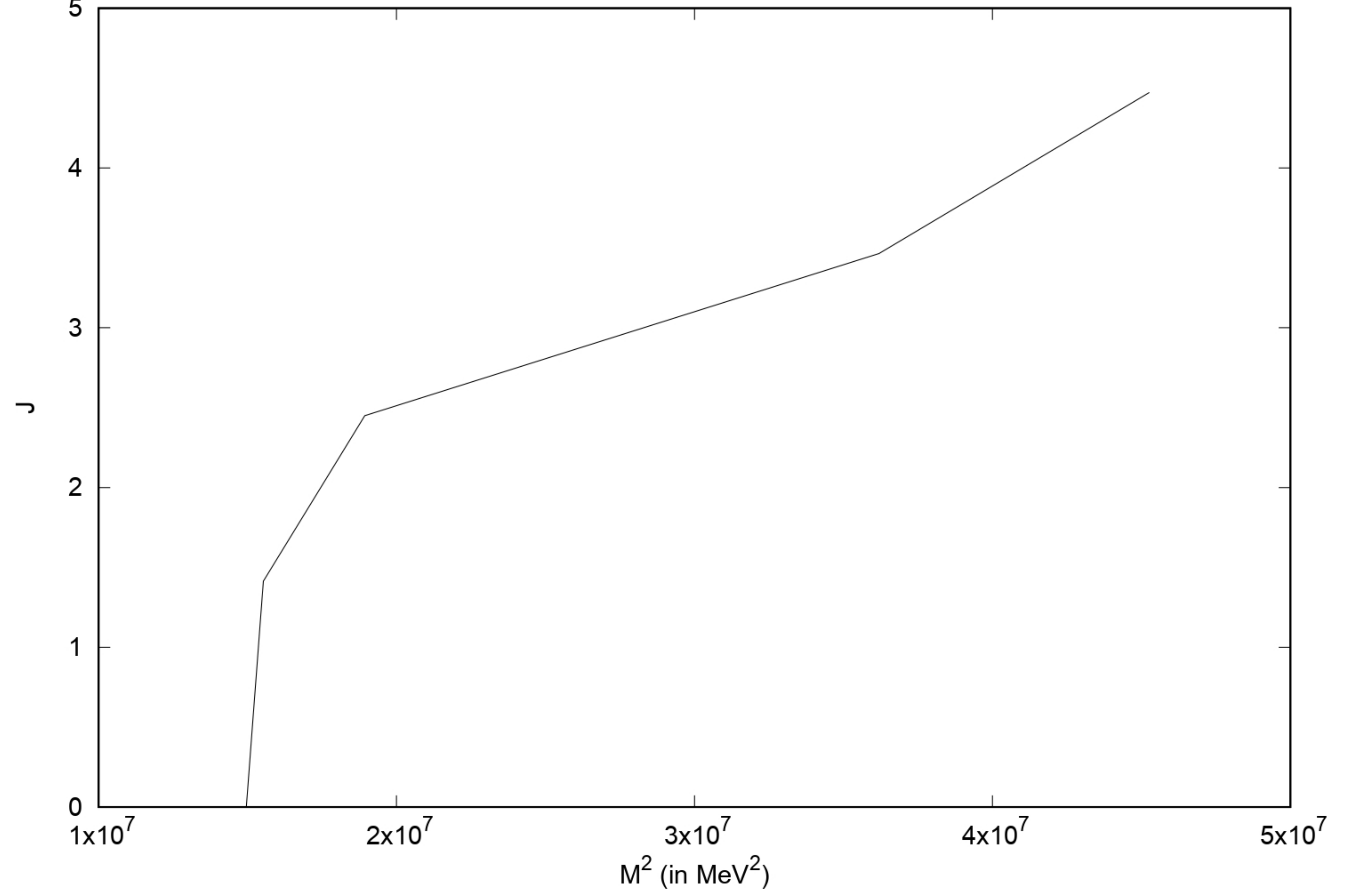}
}
\label{fig:4}       
\end{figure}

$X \longrightarrow \bar{D}^{0} D^{0}$ is one of the observed decay modes of 
$X$(3872) state \cite{patrignani}. 
These 
$\bar{D}^{0}$ and  $D^{0}$ mesons are considered as two dipoles. Using the 
energy equation, the interaction energy for $X(3872)$ tetraquark state can be 
determined. It is clear from the energy equation that the interaction energy 
depends on temperature $T$ and the reduced mass $\mu$. For $X(3872)$ 
tetraquark, the energy correction for the $s$ state with temperature $1eV$ to 
$200MeV$ ranges from $0.3MeV$ to $4000MeV$. Figure \ref{fig:2} shows that the 
mass correction increases with temperature and it also increases as we go to 
higher angular momentum state. If the heavy tetraquarks are considered, the 
energy correction is less and for light tetraquarks, the energy correction is 
more. In this framework for $X$(3872) tetraquark state the energy corrections 
and mass in different higher angular momentum states are determined and is 
mentioned in Table \ref{tab:7}.
\begin{table}[H]
\centering
\caption{The energy corrections and total mass
in different angular momentum states}
\label{tab:7} 
\begin{tabular}{ccccc}
\hline \noalign{\smallskip} 
$l$&&Mass correction&&Mass of the state\\
&&(MeV)&&(MeV)\\
\noalign{\smallskip}	
\hline \noalign{\smallskip}
$s$&&140&&3868\\
$p$&&213&&3941\\
$d$&&623&&4351\\
$f$&&2288&&6016\\
$g$&&3000&&6728\\
\noalign{\smallskip}
\hline
\end{tabular}
\end{table}
Figure \ref{fig:2} shows the mass correction for different angular momentum 
states at different temperatures (Eqn(\ref{eqn-1})). As expected for higher 
angular momentum states and temperature the mass corrections are large. This 
tells that the mass of a tetraquark also depends upon temperature of medium. 
Therefore, mass of tetraquarks produced in deconfined medium will be larger 
than the mass observed after thermal freeze out. Figure \ref{fig:3} shows the 
energy correction versus reduced mass of dipoles for different angular 
momentum states 
(Eqn(\ref{eqn-1})). The mass correction decreases with increase in reduced mass.
It indicates that the tetraquarks containing light quarks have significant 
mass contribution due to the dipole-dipole interaction. The Regge trajectory 
for the $X$(3872) system is shown in figure \ref{fig:4}. It is found to be 
highly non-linear. The same idea can be extended for Regge trajectories for 
different tetraquark systems.  
\section{Acknowledgements}
AK is thankful to Manipal Academy of Higher Education (MAHE) Manipal for the 
financial support under scheme of intramural project grant no. MAHE/CDS/PHD/IMF/2019. SDG is thankful to  `Dr. T. M. A. Pai Scholarship Program'
for the financial support.

\end{document}